\documentstyle[aps,twocolumn]{revtex}

\begin{document}
\twocolumn[\hsize\textwidth\columnwidth\hsize\csname@twocolumnfalse\endcsname

\title     
{Linear Optics Quantum Communication over Long Distances}

\author{Zeng-Bing Chen,$^1$$^*$ Huai-Xin Lu,$^1$ and 
Yong-De Zhang$^{2,1}$}
\address
{$^1$Department of Modern Physics, University of Science and Technology of China,
Hefei, Anhui 230027, People's Republic of China}
\address
{$^2$CCAST (World Laboratory), P.O. Box 8730, Beijing 100080, People's Republic 
of China}
\date{\today}
\maketitle 

\begin{abstract}
\
We propose a feasible scheme for teleporting an arbitrary polarization state or 
entanglement of photons by requiring only single-photon (SP) sources, 
simple linear optical elements and SP quantum non-demolition measurements. 
An unknown SP polarization state can be faithfully teleported 
either to a duplicate polarization state or to an entangled state. Our proposal
can be used to implement long-distance quantum 
communication in a simple way. The scheme is within the reach 
of current technology and significantly simplifies the realistc implementation
of long-distance high-fidelity quantum communication with photon qubits.

\


\

\

\end{abstract}]

Quantum mechanics offers us the surprising capabilities of manipulating
quantum information for computational and communicational tasks in a
nonclassical way \cite{QIT}. So far, many quantum information protocols have
been proposed to realize the capabilities. Among these protocols, much
attention has been paid on quantum communication (QC), e.g., quantum
cryptography \cite{BB84-Ekert} and quantum teleportation \cite
{tele-6,tele-Pan,swap-d,swap-Pan}. Quantum teleportation is a process that
transmits an unknown quantum state from a sender (Alice) to a receiver (Bob)
via a quantum channel with the help of some classical information. It may
also play a constructive role in quantum computation \cite{tele-U,Knill}. As
is widely believed, many quantum information protocols \cite{QIT}, including
QC,\ heavily rely on the pre-existing Einstein-Podolsky-Rosen (EPR)
entanglement \cite{EPR} as a necessary ``resource''. Meanwhile, quantum
entanglement is also a test ground of various fundamental problems in
quantum mechanics. Along this line of thought, a large number of physical
systems (e.g., photons \cite{Kwiat,CPZ}) has been proposed to generate
entangled states, a notoriously difficult task.

QC ultimately aims to transmit quantum states (e.g., the polarization states
of photons) or create quantum correlations between remote locations, a
long-sought goal in the context of quantum information. For transmitting an
unknown quantum state with fidelity $1$, the quantum channel is a maximally
entangled state \cite{tele-6}. Thus the establishment of high-quality
entangled states between distant nodes is essential for large-scale QC. On
the one hand, photons still remain the best long-distance carriers of
quantum information. On the other hand, the unavoidable noise of the
photonic channel degrades the entanglement established between two distant
parties. The degradation of quantum entanglement (as well as the
communication fidelity) is exponential with respect to the channel length 
\cite{repeater} and remains one of the main obstacles for robust
long-distance QC. To face this difficulty, various entanglement purification
protocol have been proposed \cite{pure,Pan-pure}. However, since
entanglement purification, though interesting in its own right, requires an
exponentially large number of partially entangled states to obtain one
desired entangled state, the quantum repeater protocol is necessary to
overcome the exponential decay of the entanglement \cite{Pan-pure}.

Motivated by several linear optics implementations of quantum information
processing (quantum computation \cite{Knill,prob-logic}, entanglement
purification \cite{Pan-pure} and generation of polarization-entangled
photons \cite{CPZ}), this paper demonstrates a dramatically different
approach for QC and quantum network over long distances by thoroughly
exploiting linear optics, as envisioned in Refs. \cite{Knill,CPZ}. We first
propose a scheme for teleporting an arbitrary polarization state or
entanglement of photons. It requires only the single-photon (SP) sources,
simple linear optical elements [e.g., the polarizing beam splitters (PBS)
and wave plates] and quantum non-demolition measurements (QNDM) with SP
resolution \cite{trigger,Kok}. An unknown SP polarization state can be
teleported either to another photon with the identical polarization state or
to an entangled state; both of the possibilities occur with the probability
of $50\%$. We then show that our scheme, together with the concept of
quantum repeaters \cite{repeater}, could be used to implement long-distance
QC in a simple way. For teleporting with certainty, one requires the
complete Bell-state measurement (BSM). Since the first teleportation
experiment, which realized the identification of only one, among four, Bell
state \cite{tele-Pan}, several feasible schemes have been proposed to
achieve the full BSM \cite{Fock-filter,Kerr,Shih}. Combining these BSM
schemes and the recent achievement of nearly ideal SP source \cite{SP-source}%
, our long-distance QC scheme is within the reach of current technology.

Very recently, Duan {\it et al}. \cite{Duan} proposed a different scheme of
long-distance QC, which involves laser manipulation of atomic ensembles,
beam splitters and SP detectors. While their proposal represents a novel way
of creating long-distance quantum entanglement between two atomic ensembles,
we are aiming to establish remote quantum entanglement between photons by
linear optics.

In the original proposal of quantum teleportation \cite{tele-6}, Alice
possesses a polarization state of photon $1$ 
\begin{equation}
\left| \psi \right\rangle _1=\alpha \left| H\right\rangle _1+\beta \left|
V\right\rangle _1,  \label{one}
\end{equation}
where $H$ ($V$) denotes the horizontal (vertical) linear polarization; $%
\alpha $ and $\beta $ are two arbitrary complex probability amplitudes
satisfying $\left| \alpha \right| ^2+\left| \beta \right| ^2=1$. The state $%
\left| \psi \right\rangle _1$, completely unknown to Alice, is to be
transmitted by Alice to Bob via one of the following quantum channels (the
Bell states) 
\begin{eqnarray}
\left| \Psi ^{\pm }\right\rangle _{23} &=&\frac 1{\sqrt{2}}(\left|
H\right\rangle _2\left| V\right\rangle _3\pm \left| V\right\rangle _2\left|
H\right\rangle _3),  \nonumber \\
\left| \Phi ^{\pm }\right\rangle _{23} &=&\frac 1{\sqrt{2}}(\left|
H\right\rangle _2\left| H\right\rangle _3\pm \left| V\right\rangle _2\left|
V\right\rangle _3),  \label{Bell}
\end{eqnarray}
which can be generated in the process of spontaneous parametric
down-conversion in a nonlinear optical crystal \cite{tele-Pan,Kwiat}. In the
four Bell states $\left| \Psi ^{\pm }\right\rangle _{23}$ and $\left| \Phi
^{\pm }\right\rangle _{23}$, the subscripts label the photon $2$ (possessed
by Alice) and the photon $3$ (possessed by Bob). Photons $2$ and $3$ (the
entangled qubits) are maximally entangled when they are in these Bell
states. In the original quantum teleportation protocol, one requires an
existing Bell state as a prerequisite resource. Currently, the mostly used
reliable source of the polarization-entangled photons is parametric
down-conversion in a nonlinear optical crystal \cite{Kwiat}. However, the
process is spontaneous (or random) and suffers from the low yield. This
limits the utility of such an entangled photon source, in particular in
long-distance QC, where the resource overheads are rather large.

Here we suggest a different quantum teleportation protocol by exploiting
only simple linear optical elements and SP sources, accompanied with
appropriate quantum measurements. Consider three independent SP sources
emitting three photons: the SP source-$1$ emits photon $1$ with quantum
state $\left| \psi \right\rangle _1$, while the remaining two SP sources
emit photons $2$ and $3$, whose quantum states are 
\begin{equation}
\left| \psi \right\rangle _j=\frac 1{\sqrt{2}}(\left| H\right\rangle
_j+\left| V\right\rangle _j).\;\;\;(j=2,3)  \label{two}
\end{equation}
These SP states in Eqs. (\ref{one}) and (\ref{two}) can be easily generated
with high precision since only trivial single-qubit operations are involved.
The current technology has realized deterministic and controllable SP
sources with high repetition rate and without the requirement of a cryogenic
temperature \cite{SP-source}, making them very suitable for the
long-distance QC task.

Now photons $2$ and $3$ are superimposed onto the PBS$_{23}$ simultaneously,
resulting in two output $2^{\prime }$ and $3^{\prime }$. Then the photon(s)
in the output $2^{\prime }$ and photon $1$ are incident on the PBS$%
_{12^{\prime }}$, with the corresponding output $1^{\prime }$ and $2^{\prime
\prime }$. Since the PBS reflects $V$-photons and transmits $H$-photons, the
whole state $\left| \psi \right\rangle _1\left| \psi \right\rangle _2\left|
\psi \right\rangle _3$ of the three photons after the action of the two PBS
is transformed into 
\begin{eqnarray}
\left| \psi \right\rangle _{1^{\prime }2^{\prime \prime }3^{\prime }} &=&%
\frac 12[\alpha \left| H\right\rangle _{1^{\prime }}\left| H\right\rangle
_{2^{\prime \prime }}\left| H\right\rangle _{3^{\prime }}+\beta \left|
V\right\rangle _{1^{\prime }}\left| V\right\rangle _{2^{\prime \prime
}}\left| V\right\rangle _{3^{\prime }}  \nonumber \\
&&+\alpha \left| 0\right\rangle _{1^{\prime }}\left| HV\right\rangle
_{2^{\prime \prime }}\left| V\right\rangle _{3^{\prime }}+\beta \left|
HV\right\rangle _{1^{\prime }}\left| 0\right\rangle _{2^{\prime \prime
}}\left| H\right\rangle _{3^{\prime }}  \nonumber  \label{PBS} \\
&&+(\alpha \left| V\right\rangle _{1^{\prime }}\left| HV\right\rangle
_{2^{\prime \prime }}+\beta \left| HV\right\rangle _{1^{\prime }}\left|
V\right\rangle _{2^{\prime \prime }})\left| 0\right\rangle _{3^{\prime }} 
\nonumber  \label{PBS} \\
&&+(\alpha \left| 0\right\rangle _{1^{\prime }}\left| H\right\rangle
_{2^{\prime \prime }}+\beta \left| V\right\rangle _{1^{\prime }}\left|
0\right\rangle _{2^{\prime \prime }})\left| HV\right\rangle _{3^{\prime }}],
\label{PBS}
\end{eqnarray}
where $\left| 0\right\rangle $ and $\left| HV\right\rangle $ denote the
zero-photon and two-photon (one with $H$-polarization and another with $V$%
-polarization) states, respectively. Introduce two new Bell states \cite
{Fock-filter} 
\begin{equation}
\left| \Xi ^{\pm }\right\rangle =\frac 1{\sqrt{2}}(\left| 0\right\rangle
\left| HV\right\rangle \pm \left| HV\right\rangle \left| 0\right\rangle ),
\label{Bell02}
\end{equation}
which are also maximally entangled. Then using Eqs. (\ref{Bell}) and (\ref
{Bell02}) enables us to rewrite the first two lines of (\ref{PBS}) as 
\begin{eqnarray}
\left| \psi ^{\prime }\right\rangle _{1^{\prime }2^{\prime \prime }3^{\prime
}} &=&\frac 12[\left| \Phi ^{+}\right\rangle _{1^{\prime }2^{\prime \prime
}}(\alpha \left| V\right\rangle _{3^{\prime }}+\beta \left| H\right\rangle
_{3^{\prime }})  \nonumber \\
&&+\left| \Phi ^{-}\right\rangle _{1^{\prime }2^{\prime \prime }}(\alpha
\left| V\right\rangle _{3^{\prime }}-\beta \left| H\right\rangle _{3^{\prime
}})  \nonumber \\
&&+\left| \Xi ^{+}\right\rangle _{1^{\prime }2^{\prime \prime }}(\alpha
\left| H\right\rangle _{3^{\prime }}+\beta \left| V\right\rangle _{3^{\prime
}})  \nonumber \\
&&+\left| \Xi ^{-}\right\rangle _{1^{\prime }2^{\prime \prime }}(\alpha
\left| H\right\rangle _{3^{\prime }}-\beta \left| V\right\rangle _{3^{\prime
}})].  \label{initial}
\end{eqnarray}

Now Bob, who possesses the output photon(s) $3^{\prime }$, needs to perform
a QNDM of the photon number at his side. His detection results will be
informed to Alice, who possesses the output photons $1^{\prime }$ and $%
2^{\prime \prime }$, via a classical channel. If the detected photon number
is zero (two), then the form of (\ref{PBS}) and the standard theory of
quantum measurement imply that Alice's states are projected onto (i.e.,
teleported to) an entangled state $\alpha \left| V\right\rangle _{1^{\prime
}}\left| HV\right\rangle _{2^{\prime \prime }}+\beta \left| HV\right\rangle
_{1^{\prime }}\left| V\right\rangle _{2^{\prime \prime }}$ ($\alpha \left|
0\right\rangle _{1^{\prime }}\left| H\right\rangle _{2^{\prime \prime
}}+\beta \left| V\right\rangle _{1^{\prime }}\left| 0\right\rangle
_{2^{\prime \prime }}$), whose entanglement is determined by the input state 
$\left| \psi \right\rangle _1$. In this case, Alice obtains an entangled
state conditioned on Bob's measurement results. However, if the photon
number detected by Bob is one, then the whole state $\left| \psi
\right\rangle _{1^{\prime }2^{\prime \prime }3^{\prime }}$ will be projected
onto $\left| \psi ^{\prime }\right\rangle _{1^{\prime }2^{\prime \prime
}3^{\prime }}$. In this case, Alice can teleport the unknown input state $%
\left| \psi \right\rangle _1$ to Bob by performing the BSM that collapses
her states onto one of the four Bell states $\left| \Phi ^{\pm
}\right\rangle _{1^{\prime }2^{\prime \prime }}$ and $\left| \Xi ^{\pm
}\right\rangle _{1^{\prime }2^{\prime \prime }}$, instead of the four Bell
states in Eq. (\ref{Bell}) in the original proposal \cite{tele-6}. The
present BSM can then be accomplished by exploiting two complete BSM schemes
proposed recently \cite{Fock-filter,Kerr}.

For convenience, we call the whole setups (the PBS, the SP sources as well
as the required measurement setups) that teleport an unknown input quantum
state a ``quantum teleporter''. The quantum teleporter can faithfully
transmit an unknown SP polarization state either to another photon with the
identical polarization state or to an entangled state; both of the
possibilities occur with the probability of $50\%$. If the latter is
discarded, the success probability of the scheme is reduced to $50\%$. Since
the perfection of the simple linear optical elements is extremely high, the
quantum teleporters in our scheme are robust and can transmit the
polarization states of photons with high fidelity.

Entanglement swapping is in fact teleportation of entanglement \cite
{tele-6,swap-d,swap-Pan}. The above scheme can also be used in entanglement
swapping. In this case we consider two EPR pairs $\left| \Psi
^{-}\right\rangle _{12}$ and $\left| \Psi ^{-}\right\rangle _{34}$; photons $%
1$ and $3$ are distributed to Alice. Alice superimposes her photons onto a
PBS ( with the corresponding output $1^{\prime }$ and $3^{\prime }$).
Afterward, the initial state $\left| \Psi ^{-}\right\rangle _{12}\left| \Psi
^{-}\right\rangle _{34}$ becomes 
\begin{eqnarray}
\left| \psi \right\rangle _{1^{\prime }23^{\prime }4} &=&\frac 12[\left|
\Psi ^{+}\right\rangle _{1^{\prime }3^{\prime }}\left| \Phi
^{+}\right\rangle _{24}-\left| \Psi ^{-}\right\rangle _{1^{\prime }3^{\prime
}}\left| \Phi ^{-}\right\rangle _{24}  \nonumber \\
&&\ \ -\left| \Xi ^{+}\right\rangle _{1^{\prime }3^{\prime }}\left| \Psi
^{+}\right\rangle _{24}-\left| \Xi ^{-}\right\rangle _{1^{\prime }3^{\prime
}}\left| \Psi ^{-}\right\rangle _{24}].  \label{swap}
\end{eqnarray}
Then Alice performs the BSM. At the end of the entanglement swapping,
conditioned on Alice's result photons $2$ and $4$ are entangled in one of
the four Bell states $\left| \Psi ^{\pm }\right\rangle _{24}$ and $\left|
\Phi ^{\pm }\right\rangle _{24}$, though they never interact in any way \cite
{swap-d,swap-Pan}. In the above entanglement swapping protocol, the two EPR
pairs can be created by the SP sources, simple linear optical elements and
the QNDM with SP resolution \cite{CPZ}. The whole setups that swap the
quantum entanglement are called an ``entanglement swapper''.

As in all realistic teleportation experiments \cite{tele-Pan}, the BSM
relies on the interference of two independently created photons. One thus
has to guarantee good spatial and temporal overlap at the PBS and above all,
one has to erase all kinds of the which-way information for the two photons.
The procedure is crucial to the validity of Eqs. (\ref{PBS}) and (\ref{swap}%
). Thus, SP sources with good coherence and controllable emission time and
direction are highly desirable and in fact, can be realized in quantum dots
with integrated cavity structure \cite{SP-source}. The SP coherence may also
be enhanced by frequency filters.

While the proposed linear optics implementation of quantum teleportation is
interesting in its own right, its usefulness lies in realizing long-distance
QC simply by combining the basic idea of quantum repeaters \cite{repeater}
and the linear optics entanglement purification protocol \cite{Pan-pure}. To
achieve a successful QC, one can either send directly the qubit, whose
information is encoded in the polarization state of a photon, through a
photonic channel (the scheme-I) or create the high-quality entanglement over
the channel, i.e., the quantum channel (the scheme-II), which can then be
used to transmit quantum information by virtue of the quantum teleportation
protocol. The quantum repeater protocol \cite{repeater} can be implemented
correspondingly in two distinct ways. In the present long-distance QC model,
one divides the whole QC channel into many nodes; the length between two
nearest neighboring nodes is comparable to the channel attenuation length.
To faithfully transmit an unknown SP polarization state, a quantum
teleporter (together with a quantum memory if needed) is used at each node:
The quantum states are transmitted by the quantum teleporters as a relay
race. Similarly, entanglement swapping can be extended to a long distance by
using many entanglement swappers, each of which is placed at one node. Most
importantly, the suggested scheme has obvious advantages in terms of its
simplicity and integrability into the linear optics entanglement
purification protocol proposed by Pan {\it et al}. \cite{Pan-pure}. The
integrability implies that the linear optics realization of long-distance QC
may also have the built-in entanglement purification, an important feature
that is essential to the long-distance QC protocol due to Duan {\it et al}. 
\cite{Duan}. The crucial point here is that our scheme of quantum
teleportation eliminates the stringent requirement of entangled photon
sources; both the entanglement and its purification are plug-in on demand.
Thus we can overcome the problem associated with the exponential decay of
the entanglement and communication fidelity in a simple way.

Another basic building block in the long-distance QC is quantum memory. It
is essential for locally storing and manipulating quantum information. As
proposed in Ref. \cite{network}, a quantum network or a QC channel consists
of spatially separated nodes, which are connected by photonic channels. When
quantum memory located at each node is represented by the internal states of
atoms (or ions), one needs to transfer the internal state of an atom at a
node to quantum state of an optical mode, and vice versa at the next node.
While this proposal is attractive for creating long-distance quantum
correlations of the internal states of atoms over a noisy channel, it is
preferable in our scheme to store the quantum states of photons within
quantum memory directly to establish the polarization entanglement of
photons over a long photonic channel. Fortunately, a light pulse can be,
with nearly $100\%$ efficiency in principle, stored for a controllable
period of time and then released on demand in an optically dense many-atom
system \cite{store-light-eprint,EIT-Nature}, as some recent experiments
demonstrated \cite{store-light}. In particular, transfer and storage of the
polarization state of a photon are also realizable \cite{store-light-eprint}%
. Storing polarization entanglement of photons over long distances could
also be accomplished by a scheme proposed by Lloyd {\it et al}. \cite{Lloyd}%
. Thus quantum memory suitable for our scheme is feasible already.

Due to dissipation, noise and imperfect measurements, errors always occur in
a realistic QC channel. Detailed analyses of the scalability and resource
requirements within our scheme are yet to be done. However, we can make the
following remarks. In some sense, our scheme is a minimal one, which has a
great advantage of tolerating errors and lessening the required overheads.
First, only three measurements, two of which are required for the complete
BSM \cite{Fock-filter}, are involved in teleporting an unknown SP state,
with six possible outcome states to be distinguished. Meanwhile, the
complicated experimental task of creating quantum entanglement does not need
to be done separately in the present approach. Moreover, the simple
linear-optical elements are thoroughly used within our scheme. Their
extremely good performance induces only small errors that can be safely
ignored. The minimal scheme we proposed is thus essential to a scalable QC
with photons.

As in the linear optics quantum computation \cite{Knill}, dominant sources
of errors in our linear optics QC come from photon loss, phase errors and
detector inefficiency, which are well understood. These errors can be dealt
with by using the method described in Ref. \cite{Knill}. The detector
inefficiency only influences the communication efficiency of our scheme,
similarly to other linear optics implementations of quantum information
processing \cite{Knill,Pan-pure}. Under the scheme-I, complicated error
detection and correction are necessarily involved \cite{QIT}. In this
respect, it is more advantageous to work under the scheme-II. Within our
linear optics realization of the quantum repeaters, the resources of
creating high-quality entanglement over a long QC channel might grow
polynomially with the channel length as in the original quantum repeater
protocol \cite{repeater}.

The QNDM with SP resolution is crucial in our approach. Standard realization
of the QNDM of the photon number requires the cross-Kerr nonlinearity \cite
{trigger}, which induces the cross-Kerr phase modulation required to
implement the complete BSM \cite{Fock-filter,Kerr}. To obtain good signal to
noise ratio, it is desirable to acquire giant cross-Kerr nonlinearity.
Though the required nonlinearity could be obtained by ultraslow SP pulses
propagating in a coherently prepared atomic gas \cite{EIT-Nature,Lukin},
currently an SP-QNDM still remains an experimental challenge. In Ref. \cite
{SP-CQED}, the SP-QNDM has indeed been realized by a cavity-QED technique.
But such an SP-QNDM is strongly frequency-dependent. Very recently, Kok {\it %
et al}. \cite{Kok} proposed a probabilistic interferometric SP-QNDM device
with linear optics and SP detectors; the device, which can distinguish the
zero-photon, single-photon and two-photon evens, is frequency-independent
and may be polarization-preserving. Thus it can be naturally integrated into
the present QC protocol.

To summarize, we have proposed a linear optics model for long-distance
high-fidelity QC. Our proposal eliminates an obstacle of the large-scale QC
since the high-quality entanglement required for faithful quantum
teleportation and entanglement swapping is plug-in on demand. It relies on
the beautiful technologies developed for nearly ideal SP sources aiming at
the QC tasks, for the QNDM with SP resolution and for quantum memory of
light. The proposed scheme simplifies the experimental aspect of realistic
long-distance QC and the optical quantum network. The linear optics quantum
information processing has several attractive advantages \cite{Knill}. With
several involved technologies working together, our scheme might be
realizable in a near future.

We thank Lu-Ming Duan and Jian-Wei Pan for valuable comments. This work was
supported by the National Natural Science Foundation of China under Grants
No. 10104014, No. 19975043 and No. 10028406 and by the Chinese Academy of
Sciences.

\end{document}